\documentclass[12pt]{article}
\def\beq{\begin{equation}}
\def\eeq{\end{equation}}
\def\barr{\begin{array}}
\def\earr{\end{array}}
\def\dis{\displaystyle}

\def\lra{\longrightarrow}
\newcommand{\lsim}
{{\;\raise0.3ex\hbox{$<$\kern-0.75em\raise-1.1ex\hbox{$\sim$}}\;}}
\def\fcap{{\;\raise1.8ex\hbox{{\tiny $\cap$}\kern-0.7em\raise-1.8ex\hbox{$f$}}}}
\def\fdc{{\;\raise2.0ex\hbox{{\footnotesize $\cap$}\kern-0.8em\raise-2.0ex\hbox{$\fcap$}}}}

\begin{document}
\thispagestyle{empty}

\begin{center}
{\Large\bf Cosmology and large mass hierarchy in multiply warped braneworld scenario}\\[20mm]
Narayan Banerjee\footnote{E-mail: narayan@iiserkol.ac.in}\\
{\em Indian Institute for Science Education and Research, Kolkata
\\Mohanpur Campus
\\Mohanpur - 741252, Nadia\\ West Bengal, India} \\[8mm]

Sayantani Lahiri\footnote{E-mail: sayantani.lahiri@gmail.com}\\
{\em Relativity and Cosmology Centre 
\\ Department of Physics}
\\{\em Jadavpur University
\\ Raja Subodh Chandra Mullick Road
\\Jadavpur
\\Kolkata- 700032, India}  \\ [8mm]

Soumitra SenGupta\footnote{E-mail: tpssg@iacs.res.in} 
\\{\em Department of Theoretical Physics\\ Indian Association for the
Cultivation of Science
\\ 2A and 2B Raja Subodh Chandra Mullick Road, Jadavpur
\\Kolkata - 700 032, India} \\

\end{center}

\bigskip
\abstract{Hubble expansion in warped braneworld model is addressed in presence of more than one warped extra dimensions. It is shown that while 
the expansion depends on all the moduli, an exponential nature of the expansion of the scale factor emerges as a generic feature
which is independent of the number of extra dimensions. Expression for the effective brane cosmological constant in such model has been derived. 
It is shown that similar to the 5-dimensional Randall-Sundrum model a fine tuning between the bulk cosmological constant and brane tension 
is required to obtain the desired cosmological constant on the brane. The length of the extra dimensions are determined in such scenario. 
Finally introduing pressureless matter in the bulk an observationally  consistent cosmology was obtained on the visible brane.}
\newpage
\section{Introduction}
The standard model of elementary particles, despite its spectacular success in explaining Physics up to TeV scale, bears an unsatisfactory feature
of unnatural fine tuning problem in connection with the large radiative correction of mass of the Higgs scalar \cite{Dress}. 
Extra dimensional models \cite{arkani,randall,rubakov} successfully resolved this problem and at the same time invoked new Physics beyond the 
Standard model \cite{antoniadis,witten,lykken,cohen}. The warped geometry model, proposed by Randall and Sundrum \cite{randall}, is one of the 
significant steps towards this direction. Various phenomenological as well as cosmological implications of this model have been studied extensively 
during the past decade and many interesting predictions have been made which are to be tested in the forthcoming TeV scale and cosmological experiments. 
While the original Randall Sundrum model was formulated with one warped dimension, a generalisation of this model was proposed in the 
presence of more than one warped dimensions \cite{rshigh}. Such models give rise to additional new features which may be observed in the future 
collider experiments\cite{csaki}. While some of the cosmological implications have already been studied in the context of 5-dimensional 
RS model \cite{lang,kaloper,nihei,kolda,bum,lim,sahni}, an extensive investigation in the context of multiply warped model is yet to be done. In this paper 
we propose to study the features of Hubble expansion of the universe in the context of multiply warped model and 
explore new Physics at the cosmological scale to search for extra dimensions. \\
We begin with a brief description of the RS model and it's cosmological significance. Then we describe the multiply warped model to find the 
role of additional warped dimensions in respect to the expanding model of the universe.

\section{Five dimensional Randall-Sundrum Model}
     \label{sec:6d}

Randall-Sundrum model \cite{randall} considers a five dimensional anti de-Sitter spacetime where the extra co-ordinate $y$ is compactified on a $S^1/Z_2$ manifold. Two 3-branes are located at the orbifold fixed points $y=0$ and $y=\pi$. 
The action for this model is given by :
\beq
\barr{rcl}
S & = & \dis S_{gravity} + S_{pl} + S_{vis}   \,, \\[1ex]

S_{gravity} & = & \dis \int {d^4 x} \, \int\limits_{-\pi}^{\pi}{d y}  \, 
          \sqrt{-g_5} \; \left(\frac{M_*^3}{2}\, R_5 - \Lambda_5 \right) 
\\[2ex]
S_{pl} & = & \dis \int d^4 x \sqrt{-g_{pl}}[{\cal L}_{1} - \lambda_{1}] 
 .\\[2ex]
 
S_{vis} & = & \dis \int d^4 x \sqrt{-g_{vis}}[{\cal L}_{2} - \lambda _{2}]            \label{eq:action}
\earr
\eeq

\noindent where $ {\cal L}_{1}, {\cal L}_{2}  $ are Lagrangian on the Planck brane and visible brane respectively while $ \lambda_{1}, \lambda _{2} $ are corresponding brane tensions. The $ 5 $-dimensional bulk contains only the cosmological constant $ \Lambda_5 $ and $ M_* $ is the $ 5 $D Planck mass scale.

\noindent The geometry of the 5D spacetime is $[M^{1,3} \times S^1/Z_2]$ with a metric ansatz,
\beq
ds^2 = e^{-2\sigma(y)}\eta_{\mu\nu}dx^{\mu}dx^{\nu} + b_0^2 dy^2 .
    \label{metric} 
\eeq 
Here $\eta_{\mu\nu} = diag(-1,1,1,1)$ is the four dimensional Minkowski metric. $b_0$ is the radius of $S^1$ and $y$ is an angular co-ordinate running from $0$ to $\pi$. Therefore the proper length of the extra dimension is 
{$L_y=b_0\pi$.}\\
Solution of the 5D Einstein's equation corresponding to this metric incorporating $ Z_2 $ symmetry, yields
\begin{center}
\beq
\sigma = b_0|y|k             \label{sig}
\eeq
\end{center}
\noindent where $k = \sqrt{\dis \frac{-\Lambda_5}{6M^3}}$\quad.\\

\noindent In the RS scenario, the relation between bulk cosmological constant $\Lambda_5$ and brane tensions $\lambda_1$ and $\lambda_2$ is as follows :\\
\beq
k = k_1 = -k_2                                                     \label{eq:12}
\eeq
\noindent {where $k =\sqrt{\dis \frac{-\Lambda_5}{6M^3}}$ , $k_i = \dis \frac{ \lambda_i}{6M^3}$ \quad $(i\,=\,1,2 )$\\ [1ex]

\noindent The solution given by eqn.({\ref{sig}}) suggests that the bulk is five dimensional AdS spacetime with negative bulk cosmological constant $\Lambda _5 $ such that the warped solution is obtained. The 3-brane situated at $y=0$ suffers no warping and is called the hidden brane while the Standard model 3-brane located at $y=\pi$, is exponentially warped. The brane tensions on the two 3-branes as seen from eqn.({\ref{eq:12}}) are identical in magnitude but the brane tension turns out to be positive on the hidden brane and negative on the Standard Model brane. Subsequently it was shown that the brane separation modulus $ b_0 $  can be stabilized by introducing a bulk scalar field \cite{gw} without further unnatural fine-tuning.
The effective $4D$ visible brane cosmological constant is \cite{shiromizu}
\beq
\Lambda_4^{eff}\,=\,\frac{1}{2}\,\kappa_5^2\,\dis\left(\Lambda_5\,+\,\frac{1}{6}\,\kappa_5^2\,\lambda^2_2\right) \quad.          \label{eq:cos_eff}
\eeq
\noindent where $ \kappa_5 $ is related to the $ 5 $-dimensional gravitational constant and $\lambda_2 $ is the brane tension of the four dimensional visible world. Using the expression of the brane tension $ \lambda_{i} $, the induced cosmological constant on both the branes in the Randall-Sundrum model turns out to be zero thereby making each of them flat which can be observed particularly for visible brane from eqn.({\ref{eq:cos_eff}}) using the expression for $ k_2 $. Due to warped geometry, the physical mass on the visible brane suffers an exponential warping,
\begin{center}
\beq
 m = e^{-kb_0\pi}m_0 
 \eeq
\end{center}
Thus mass of Higgs scalar is always exponentially warped on the Standard Model brane. With $k b_0\,\simeq\,11.5$ one can achieve the desired Planck to TeV scale warping without introducing any unnatural fine tuning. In this model the four dimensional Planck mass scale $ M_{pl} $ is related to five dimensional Planck scale as,
\begin{center}
\beq
M_{pl}^2 = \frac{M_*^3}{k} (1 - e^{-2 kb_0\pi})
\eeq
\end{center}

\section{Inflation in five dimensional spacetime } \label{sec:5Dinflation}
We briefly discuss how inflation \cite{kim} in our Universe can arise from its embedding in a five dimensional warped geometry $ [M^{1,3}\times S^{1}/Z_2] $ with AdS bulk. Our Universe is a 3-brane located at $y=\pi$ while the other brane situating at  the orbifold point $y=0$ is the Planck brane. The 3-space of the visible universe is assumed to be homogeneous and isotropic and is also taken to be flat. In this five dimensional model its time evolution is governed by a scale factor, $R(y,t) = f(y)v(t)$ which smoothly converges to Randall- Sundrum model in the appropriate static limit. The action for the above set up is given by eqn.({\ref{eq:action}}). The metric ansatz for such a warped cosmological model is given by :
\beq
ds^2\,=\,f^2(y)[-dt^2 + v^2(t)\delta_{ij}dx^i dx^j] + b_0^2  dy^2 \label{eq:1}\
\eeq
where $b_0$ is a constant. 

\noindent The proper length along the direction of the compact co-ordinate $y$ is : $L_5 = b_0\pi$ . In the absence of inflation one receives the static limit of RS model given by eqn.({\ref{eq:12}}).

In this scenario we additionally introduce energy densities i.e. $\rho$ on the 3-branes. As a result from $ (tt) $ and $ (ii) $ components of Einstein's equations we find : 
\beq
v(t) = e^{H_0t}
\eeq
 where $H_0$ is the Hubble's constant. Now from $(55)$ component of Einstein's equation one can obtain the solution for \,$f(y)$\,in terms of $k$ and $H_0$ consistent with $ Z_2 $ symmetry as :
\beq
f(y) = \frac{H_0}{k}\sinh (- k b_0 |y| + d_0) \label{eq:2}\
\eeq
where $ d_0 $ is the constant of integration.
\\
\noindent The resulting cosmological warped metric therefore becomes :
\beq
ds^2 =\left(\frac{H_0}{k}\right)^2\sinh^2 (- k b_0 |y| + d_0)[-dt^2 + e^{2H_0t}\delta_{ij}\,dx^i dx^j] +\, b_0^2\,  dy^2        \label{metric_1}
\eeq
This metric describes inflation of the spatial three dimensions. In the static limit when $H_0\lra 0$ and $d_0\lra \infty$ with the ratio $(\frac{H_0}{2k})^2 e^{2d_0}\lra 1$, the metric coefficient becomes that of static RS metric.

\noindent Using the boundary conditions, the brane tensions on the two 3-branes are :
\beq
k_1 = k\coth (d_0)             \label{eq:k1}
\eeq
\beq
-k_2 = k \coth {( -k b_0 \pi  + d_0)}         \label{eq:k2}
\eeq
which also reduce to eqn.({\ref{eq:12}}) in the static limit.
\\
\noindent In the non-static scenario, the effective $4D$ cosmological constant on the visible brane is now non-zero. However in the static limit when $ H_0 = 0 $ we retrieve $ \Lambda_4^{eff} = 0 $ [using eqn.(\ref{eq:cos_eff})]. Moreover, the non-static solutions given by eqn.({\ref{eq:k1}}) and eqn.({\ref{eq:k2}}), are satisfied only when $k<k_1<-k_2$.

\noindent Now once again using eqn.({\ref{eq:k1}}) and eqn.(\ref{eq:k2}) the length of the extra dimension $L_5$ can be written in terms of $ k_1, k_2$ and $ k $ . The resulting expression for $ L_5 $ turns out to be :
\beq
L_5 = b_0\pi = \frac{1}{2k}\ln \left[\frac{-k_2 - k}{k_1 - k}\frac{k_1 + k}{-k_2 + k}\right]
\eeq
\\
As we can see from the above equation, in the inflationary Universe the length of the extra dimension is kept fixed, which in turn requires fine tuning between bulk cosmological constant and brane tensions. This fine-tuning feature also exists in the original RS model that produces a vanishing effective cosmological constant on the 3-branes. However, it may be noted that in the static limit one cannot express $ L_5 $ in terms of $ \Lambda_5 $ and $ \lambda_i $'s.

\noindent Now the effective four dimensional metric can be obtained  from six-dimensional metric eqn.({\ref{metric_1}}) by an appropriate co-ordinate transformation which becomes :
\beq
 ds^2_{4} = -dt^2 + e^{2H(y)t}\delta_{ij}dx^i dx^j
\eeq
where Hubble parameter is now given as :
\beq
H(y) = k\,cosech (-k b_0 |y| + d_0)
\eeq
{At the boundaries, i.e. on $y=0$ and $y=\pi$ orbifold fixed points, the values of Hubble parameter are :}\\
{ $H(0)   = \sqrt{k_1^2 - k^2}$}\\
{ $H(\pi) = \sqrt{k_2^2 - k^2}$}\\

\noindent On the Standard model brane, when our Universe is evolving with time, its non-zero Hubble parameter is expressed as fine tuning between bulk cosmological constant $ \Lambda_5 $ and visible brane tension $ \lambda_2 $  given by the relation,
\beq
H(\pi) = \sqrt{k_2^2 - k^2}{\lsim {10^{-61 }}}M_{pl}
\eeq
It is easy to show that the static limit, $ k_2 = k $ yields $ H(\pi) = 0 $. The size of the extra dimension can be written as the ratio of Hubble parameters. In this model the problem related to fine-tuning of Higgs mass is resolved at the expense of more severe fine tuning between bulk cosmological constant and brane tension of the Planck brane.
\beq
\barr{c}
{kL_{RS} = kb_{RS}{\pi}\simeq{\ln \frac{H(\pi)}{H(0)}}}\\

H(0) = \sqrt{k_1^2 - k^2}\times {10^{-16}\
\Rightarrow H(0)}\lsim{10^{-77}M_{pl}}
\earr
\eeq
We now examine all the issues related to inflation in warped geometry in the presence of a multiply warped spacetime which results from a natural generalisation of the five dimensional Randall-Sundrum model.

\section{Six-dimensional doubly warped geometry}
In order to generalise the Randall-Sundrum model to six dimensions \cite{Multiply} one can consider the doubly warped  spacetime as $M^{1,5} \rightarrow [M^{1,3} \times S^1/Z_2] \times S^1/Z_2$ with AdS $(\Lambda_B < 0)$ bulk spacetime. The non-compact co-ordinates $x^\mu$ run over usual four dimensional spacetime while $y$\,,\,$z$ are compactified angular co-ordinates. Such a geometry gives rise to a brane-box like spacetime in which the 4-branes are placed at the orbifold fixed points namely at $y = 0,\pi$ and $ z = 0,\pi$. There are four 3-branes which exist at the intersection of two 4-branes.

\medskip
\underline{Metric ansatz :}\\
\begin{center}
{$ds^2 = b^2(z)[ a^2(y) \eta_{\mu\nu} dx^{\mu} dx^{\nu} + b_0^2  dy^2] + c_0^2  dz^2$}\\ \label{eq:3}
\end{center}

The total bulk-brane action is given by,
\beq
\barr{rcl}
S & = & \dis S_6 + S_5 + S_4 \\[1ex]
S_6 & = & \dis \int {d^4 x} \, {d y} \, {d z} \, 
          \sqrt{-g_6} \; \left(\frac{M^4}{2}R_6 - \Lambda_B \right) 
\\[2ex]
S_5 & =  & \dis -\int {d^4 x} \, {d y} \, {d z} \, 
       \left[ \lambda_1 \, \delta(y) + \lambda_2 \, \delta( y - \pi) \right]\sqrt{-g_5}
\\[1.5ex]
    & - & \dis \int {d^4 x} \, {d y} \, {d z} \, 
         \left[ \lambda_3 \, \delta(z) + \lambda_4 \, \delta(z - \pi) \right]\sqrt{-\tilde{g}_5}
\\[2ex]
S_4 & = & \dis \int d^4 x dy dz \sqrt{-g}[{\cal L} - \lambda] \ .
\earr
    \label{Action1}
\eeq
Note that, the brane potential terms $\lambda_{1} = \lambda_{1}(z) $, $\lambda_{2} = \lambda_{2}(z)  $ whereas $\lambda_{3} = \lambda_{3}(y) $, $ \lambda_{4} = \lambda_{4}(y) $ can in principle be the functions of $y$ and $z$. The 
 term $S_4$ corresponds to the 3-branes
placed at $(y, z) = (0,0), (0, \pi), (\pi, 0), (\pi, \pi)$.
\\
On substituting the metric and solving six-dimensional Einstein's equations and further imposing $Z_2$ symmetry, the warp factors turn out to be :\\

\beq
a(y) = e^{-c\,|y|} \label{warp1}
\eeq
\
\beq
b(z) = \dis \frac{\cosh(kz)}{\cosh(k\pi)} \label{warp2}
\eeq

where
\begin{center}
\begin{normalsize}
\beq
c = \frac{b_0 k}{ c_0 \cosh(k \pi)} \label{eq:c}
\eeq 

\beq
k =  c_0 \, \sqrt{\dis \frac{-\Lambda_B}{10 M^4}}
\eeq
\end{normalsize}
\end{center}

\noindent Solving Einstein's equations the solution for the warped metric is given by :

\begin{eqnarray}
ds^2 & = & \dis \frac{\cosh^2(k  z)}{\cosh^2 (k  \pi)} \,
  \left[ \exp\left(- 2  c  |y| \right)
                \, \eta_{\mu \nu} \, d x^\mu d x^\nu 
     + b_0^2 \, d y^2 \right] + c_0^2 \, d z^2 \ .
\end{eqnarray}
\noindent Now two of the 4-brane tensions on the 4-branes $y=0$ ,\,$y=\pi$  are co-ordinate dependent and are given by : \\
\beq
\lambda_1(z)=\,-\lambda_2(z) = \, 8M^2 \,\sqrt{\dis\frac{-\Lambda_B}{10}}\,sech (kz) 
\eeq
while on 4-branes located at  $z=0$ and $z=\pi$, the brane tensions are respectively : \\
\begin{eqnarray}
\lambda_3(y)\,=\,0 \\
\lambda_4(y)\,=\,\frac{-8 M^4 k}{c_0} \tanh (k\,\pi)\
\end{eqnarray}

\noindent From these, the 3-brane placed at $y = 0$ , $z = \pi $  suffers minimum warping and is called the Planck brane with brane tension \\
\beq
\lambda_{Planck}= 8M^4\sqrt{\frac{-\Lambda_B}{10 M^4}}\:[sech(k\pi)-\tanh (k\pi)] \ 
\eeq

\noindent The visible 3-brane located at $ y=\pi $ , $ z=0 $ has brane tension,\
\beq
\lambda_{vis}=-8M^4\sqrt{\frac{-\Lambda_B}{10 M^4}}
\eeq
\noindent The Planck sale mass $m_0$ is now warped to
\beq
m = \,m_0\,\frac{e^{-c\,\pi}}{\cosh(k\pi)}
\eeq

\beq
\Rightarrow
\frac{e^{-c\,\pi}}{\cosh(k\pi)}\approx 10^{-16}
\eeq
on the visible brane.\\
It is observed from eqn.(\ref{eq:c}) that one cannot have equal amount of warping along both the compact co-ordinates without introducing large hierarchy between the two moduli $b_0$ and $c_0$.\
An important phenomenological consequence of the model is a possible explanation of the mass hierarchy of the Standard model fermions on the  brane. The Standard model fields in each of these 3-branes have mass-scales close to TeV with some splitting among them.

\section{Inflation in six-dimensional doubly warped spacetime}
In the present work, we explore  cosmological implications of a six-dimensional spacetime in which both the extra dimensions are compactified on circles with $Z_2$ orbifolding, thus giving rise to a doubly warped spacetime as discussed in the previous section \cite{Multiply}. In such a spacetime $S^1/Z_2$ orbifoldings are not independently implemented but are performed successively, so that with every warping one warped compact dimension gets further warped. One can extend the procedure incorporating several extra dimensions leading to a multiply warped spacetime.
However, we shall first study the inflation in the background of doubly warped spacetime. The manifold under consideration is :\
$M^{1,5} \rightarrow [M^{1,3} \times S^1/Z_2] \times S^1/Z_2$. \

\subsection{Set up}
We consider a $(5+1)$ dimensional AdS bulk where two successive $S^1/Z_2$ orbifoldings along the two extra dimensions lead to a brane-box like spacetime, whose walls are 4-dimensional branes. Each of the 4-branes are situated at the orbifold fixed points namely, $y=0\,,\pi$ and $z=0\,,\pi$. At the intersection region of two such 4-branes, a 3-brane is formed. Therefore four such 3-branes reside on the points:\\
 $(y\,,z) : (0\,,0)\,,\,(0\,,\pi)\,,\,(\pi\,,0)\,,\,(\pi\,,\pi)$\\ [1mm]
As we are interested to study cosmological behaviour, we consider our Universe (i.e. one of the 3-branes) to be homogeneous and isotropic and  for simplicity we take it spatially flat 3-space. In this model the time evolution of 3-universe is governed by the scale factor which naturally is a  function of both the compact co-ordinates $y\,,\,z$.\\
 
\noindent\underline{Notations} :\\
The non-compact co-ordinates : $\textit{x}^{\mu}$ where ${\mu\,=\,0,1,2,3}$.\\
Bulk indices  : $M\,,\,N$ run over $M\,,\,N\,=\,{\mu}\,,y\,,\,z$.\\ 
$\tilde{\alpha},\tilde{\beta}$ run over $x^\mu$ and $y$ while $\alpha,\beta$ run over $x^\mu$ and  $z$. \\

\noindent \underline{Cosmological metric ansatz} :\\
\beq
 ds^2 = b^2(z)[ a^2(y)\,(-dt^2\,+\,v^2(t)\,\delta_{ij}\,dx^i\,dx^j)\, + \,b_0^2\, dy^2] \,+\, c_0^2\, dz^2 \label{eq:4}
 \eeq
where $b_0$ and $c_0$ are the moduli along the compact co-ordinates $y$ and $z$ respectively which in our proposed model are taken as constants.\\ [1mm]
The scale factor can easily be identified as $R(t,y,z)\,=\,v(t)\,a(y)\,b(z)$\\ [1mm]
The six-dimensional bulk-brane action is given by \
\beq
\barr{rcl}
 S & = & \dis S_6 + S_5 + S_4 \\[1ex]
 S_6 &  =  & \dis \int {d^4 x} \, {d y} \, {d z} \, \sqrt{-g_6} \;\left(\frac{M^4}{2}R_6 - \Lambda_B \right) 
\\[2ex]
 S_5 & = & \dis \int {d^4 x} \, {d y} \, {d z} \, \left\{\sqrt{-g_5}\;[\,{\cal L}_1\,-\,\lambda_1(z)] \: {\delta(y)}\,+\,\sqrt{-g_5}\;[ \,{\cal L}_2 \,-\, \lambda_2(z)] \: {\delta( y - \pi)} \,\right\}
\\[1.5ex]
 & +  & \dis \int {d^4 x} \, {d y} \, {d z} \, \left\{\sqrt{-\tilde{g}_5}\;[\,{\acute{\cal L}}_3\,-\,\lambda_3(y)]\,{\delta(z)}\,+\,\sqrt{-\tilde{g}_5}\,[{\acute{\cal L}}_4\,-\,\lambda_4(y)]{\delta(z-{\pi})}\,]\right\}
\\[2ex]
 S_4 & =  & \dis \int d^4 x dy dz \sqrt{-g}[{\cal L} - \lambda] \ .
\earr
         \label{Action2}
\eeq

\noindent Here, $\lambda_1$ , $\lambda_2$ , $\lambda_3$ and $\lambda_4$ are co-ordinate dependent brane tensions. ${\cal L}_1 $, ${\cal L}_2 $ are the Lagrangian of matter at $y=$ constant $4$-branes, while ${\acute{\cal L}}_{3}$, ${\acute{\cal L}}_{4}$ are the Lagrangian of matter at $z=$ constant $4$-branes. $S_4$ gives the contribution of 3-branes.\\
Since we are interested in cosmology, we assume the matter content on the 4-branes as well as on 3-branes to be perfect fluids described by energy density $\rho$ and fluid pressure $p$ in addition to the presence of corresponding brane tensions as have been discussed earlier.\

\noindent By varying the action $S$ the full six dimensional Einstein's equation is obtained as,\
\beq
\barr{rcl}
- \dis M^4 \, \sqrt{-g_ 6} \, \left( R_{MN} - \, \frac{R_6}{2} \, g_{MN}\right)
& = & \dis \Lambda_{B} \, \sqrt{-g_ 6} \: g_{MN} \\
& - & \dis 
  
\dis \left[\left(T^{\gamma}_{\beta}\,g_{\alpha\gamma}\right)_1\,+\,\,\dis \left(T^{\gamma}_{\beta}\,g_{\alpha\gamma}\right)_2\, \right] \sqrt{-g_5} \,\delta^{\alpha}_{M} \, \delta^{\beta}_{N}  \\[1ex]

& - & \dis \left[\left(T^{\tilde{\gamma}}_{\tilde{\beta}}\,\widetilde g_{\tilde{\alpha} \tilde{\gamma}}\right)_3\,+\,\left(T^{\tilde{\gamma}}_{\tilde{\beta}}\,\widetilde g_{\tilde{\alpha} \tilde{\gamma}}\right)_4\right]  \sqrt{-\widetilde g_5}\,  \delta^{\tilde{\alpha}}_{M} \, \delta^{\tilde{\beta}}_{N} \\[1ex]
\earr
\eeq
$g,\widetilde{g}$ are the respective metrics in these (4+1)-dimensional subspaces.\\

\noindent \textbf{Energy-momentum tensor on the $4+1$ branes : }
 \\
\beq
(T^{\gamma}_{\beta})_1\,=\,diag\left[-\rho_1(z) - \lambda_1(z)\,,\,p_1(z) - \lambda_1(z)\,,\,p_1(z) - \lambda_1(z)\,,\,p_1(z) - \lambda_1(z)\,,0\,,\,p_1(z) - \lambda_1(z)\,\right]\delta(y) \,                \label{T_1}
\eeq

\beq
(T^{\gamma}_{\beta})_2\,=\,diag\left[-\rho_2(z) - \lambda_2(z)\,,\,p_2(z) - \lambda_2(z)\,,\,p_2(z) - \lambda_2(z)\,,\,p_2(z) - \lambda_2(z)\,,0\,,\,p_2(z) - \lambda_2(z)\,\right]\delta(y-\pi)                       \label{T_2}
\eeq

\beq
(T^{\tilde{\gamma}}_{\tilde{\beta}})_3\,=\,diag\left[-\rho_3(y) - \lambda_3(y)\,,\,p_3(z) - \lambda_3(y)\,,\,p_3(y) - \lambda_3(y)\,,\,p_3(y) - \lambda_3(y)\,,p_3(y) - \lambda_3(y)\,,\,0\,\right]\delta(z)               \label{T_3}
\eeq

\beq
(T^{\tilde{\gamma}}_{\tilde{\beta}})_4\,=\,diag\left[-\rho_4(y) - \lambda_4(y)\,,\,p_4(z) - \lambda_4(y)\,,\,p_4(y) - \lambda_4(y)\,,\,p_4(y) - \lambda_4(y)\,,p_4(y) - \lambda_4(y)\,,\,0\,\right]\delta(z-\pi)                \label{T_4}
\eeq

\noindent Substituting the metric, the various components of Einstein's equations yield : \\ 

\noindent \underline{\textit{tt component}} :\\
\beq
\barr{rcl}
-M^4 \dis \left[\frac{-3aa''}{b_0^2}\,-\,\frac{4a^2b\,\bar{\bar{b}}}{c_0^2}\,-\,\frac{3\,a'^2}{b_0^2}\,-\,\frac{6\,a^2\,\bar{b}^2}{c_0^2}\,+\,\frac{3\,\dot{v}^2}{v^2}\right]\,(b_0\,c_0)&=
 & \dis
-\Lambda_B\,(\,b^2a^2b_0c_0\,)\\
&-& \dis
\left\{(\rho_{1}\,+\,\lambda_{1})\,\delta(y)\right\}\,(ba^2c_0) \\ [1ex]
&-& \dis
\left\{(\rho_{2}\,+\,\lambda_{2})\,\delta(y-\pi)\right\}(ba^2c_0)\\ [1ex]
 &-& \dis 
 \left\{(\rho_{3}\,+\,\lambda_{3})\,\delta(z)\right\} (b^2a^2 b_0) \\[1ex]
 &-& \dis 
 \left\{(\rho_{4}\,+\,\lambda_{4})\,\delta(z-\pi)\right\} (b^2a^2 b_0)\\[2ex]
 \earr
\eeq

\noindent \underline{\textit{ii component}} :\\
\beq
\barr{rcl}
-M^4 \dis
\left[-\dot{v}^2\,+\, \frac{3 a'^2v^2}{b_0^2}\,+\,\frac{3av^2a''}{b_0^2}\,+\,\frac{6a^2\bar{b}^2v^2}{c_0^2}\,+\,\frac{4a^2v^2b\bar{\bar{b}}}{c_0^2}-2v\ddot{v}\right](b_0c_0) & = & \dis
\Lambda_B\,\,(b^2 a^2 v^2 b_0 c_0)\\
& + & \dis
\left\{(\lambda_{1}\,-\,p_{1})\,\delta(y)\right\}\,(ba^2 v^2 c_0) \\ [1ex]
&+& \dis
\left\{(\lambda_{2}\,-\,p_2)\,\delta(y-\pi)\right\}(ba^2 v^2 c_0)\\ [1ex]
 &+& \dis 
  \left\{(\lambda_{3}\,-\,p_3)\,\delta(z)\right\} (b^2a^2 v^2 b_0) \\[1ex]
&+& \dis 
  \left\{(\lambda_{4}\,-p_4)\,\delta(z-\pi)\right\} (b^2a^2 v^2 b_0) \\[2ex]

\earr
\eeq

\noindent \underline{\textit{yy component}} : \\
\beq
\barr{rcl}
\dis
-M^4\left[\frac{6b_0^2\,\bar{b}^2}{c_0^2}\,+\,\frac{4b\,b_0^2\,\bar{\bar{b}}}{c_0^2}\,+\,\frac{6a'^2}{a^2}\,-\,\frac{3b_0^2\dot{v}^2}{a^2\,v^2}\,-\,\frac{3b_0^2\ddot{v}}{a^2v}\right] c_0 & = & \dis \Lambda_B\, (b^2b_0^2c_0)\\
             & + &  \,\,\left\{(\lambda_3\,-\,p_3)\,\delta(z)\right\}(b^2\,b_0^2) \\ [1ex]
             & + & \dis 
\left\{(\lambda_{4}\,-\,p_4)\delta(z-\pi)\right\}(b^2\,b_0^2) \\[1ex]
\earr
\eeq

\noindent \underline{\textit{zz component}} : \\
\beq
\barr{rcl}
\dis
-M^4\left[\frac{-3\,c_0^2\,\dot{v}^2}{a^2\,b^2\,v^2}\,+\,\frac{10\bar{b}^2}{b^2}\,+\,\frac{6\,c_0^2\,a'^2}{b^2b_0^2 a^2}\,+\,\frac{4\,c_0^2\,a''}{b_0^2\,b^2\,a}\,-\,\frac{3 c_0^2\ddot{v}}{a^2 b^2 v}\right](b\,b_0)& = & \Lambda_B(b\,b_0\,c_0^2) \\
             &+&\,{(\lambda_{1}\,-\,p_1)\,\delta(y)}\,c_0^2 \\ [1ex]
             &+& \dis
             (\lambda_{2}\,-\,p_2)\,\delta(y-\pi)\,c_0^2\\  [2ex]
\earr
\eeq

\noindent where dots represent differentiation with respect to $t$, primes represent differentiation with respect to $y$ and bar denotes differentiation with respect to $z$.\\[1mm]

\noindent \textbf{Determination of the function $v(t)$ :}\\
From $(tt)$  and $(ii)$ components of Einstein's equation we get,
\beq
\barr{rcl}
2\,(-M^4)\,b_0\,c_0(\dot{v}^2\,-\,v\,\ddot{v}) = 0
\earr
\eeq
which on solving gives,
\beq
v(t)\,=\,Ae^{H_0t} \label{eq:5}
\eeq
where $A$ is an integration constant and $H_0$ identified as Hubble's constant.
Interestingly, the form of the solution in eqn. ({\ref{eq:5}}}) does not depend on the number of extra spatial dimensions.
\\
\\
\noindent \textbf{Determination of the warp factors $a(y)$ and $b(z)$ :}\\
To determine the warp factors, we substitute eqn. (\ref{eq:5}) in $yy$ component which reduces to,\
\beq
\barr{rcl}
\dis
(-M^4)\,c_0\left[\frac{6 b_0^2\,\bar{b}^2}{c_0^2}\,+\,\frac{4 b\,b_0^2\, \bar{\bar{b}}}{c_0^2}\,+\,\frac{6 a'^2}{a^2}\,-\,\frac{3 b_0^2\,H_0^2}{a^2}\,-\,\frac{3 b_0^2\, H_0^2}{a^2}\right] \, = \, \dis(b^2 b_0^2) \left[\Lambda_B c_0 + (\lambda_3 - p_3)\delta(z) + (\lambda_4 - p_4)\delta(z-\pi)\right] \label{eq:6}
\earr
\eeq

\noindent Starting with the bulk part of eqn.(\ref{eq:6}) and rearranging terms we get,
\beq
\barr{rcl}

\dis\left(\frac{a'^2}{a^2}\,-\,\frac{b_0^2\,H_0^2}{a^2}\right) =  d^2  = 
\dis -b_0^2\left[\frac{\bar{b}^2}{c_0^2}\,+\,\frac{2\,b\,\bar{\bar{b}}}{3\,c_0^2}\,+\,\frac{b^2\Lambda_B}{6M^4}\right]\\
\earr
\eeq
where d is an arbitrary constant.
We thus obtain two separate equations :

\beq
\barr{rcl}
\dis
\left(\frac{a'^2}{a^2}\,-\,\frac{b_0^2\,H_0^2}{a^2}\right)  = \: d^2 \label{eq:7}\\[1mm]
 \earr
\eeq

\beq
\barr{rcl}
\dis
b_0^2\left[\frac{\bar{b}^2}{c_0^2}\,+\,\frac{2\,b\,\bar{\bar{b}}}{3\,c_0^2}\,+\,\frac{b^2\Lambda_B}{6M^4}\right] & = & -d^2 \label{eq:8}\
\earr
\eeq

\noindent The solution of the eqn.(\ref{eq:7}) consistent with the $Z_2$ symmetry gives warp factor along the compact $y$ co-ordinate as :
\beq
a(y)\, = \,\frac{b_0\,H_0}{d}\,\sinh(-d\,|y|\,+\,d_0) \label{eq:9}
\eeq
where $ d_0 $ is the constant of integration.
\\
Solving eqn.(\ref{eq:8}) gives the warp factor along the $z$ co-ordinate as,
\beq
b(z)\,=\,\frac{\cosh(kz)}{\cosh(k\pi)} \label{eq:10}
\eeq

\noindent with
\beq
\barr{rcl}
\dis
d = \frac{b_0 k}{c_0 \cosh(k\pi)} \:,\qquad \label{eq:11}
k = c_0\sqrt{\frac{-\Lambda_B}{10 M^4}}
            \label{d}
\earr
\eeq

\noindent Both the solutions of $a(y)$ and $b(z)$ give the bulk  solution of the geometry. It should be noted that once again the five dimensional solution discussed in section 3 can be recovered in the limit $c_0 = 0 $ and $b(z)=1$ [see eqn. (\ref{eq:1})]. In the static limit :
$d_0\longrightarrow\infty$ and $\dis \frac{H_0\,\cosh(k\pi)}{\sqrt{\frac{-\Lambda_B}{10 M^4}}}\,e^{d_0}\longrightarrow 1$ , one obtains the warp factor $a(y)$ as in eqn.(\ref{warp1}). It may be noted that the functional form of $b(z)$ is independent of time  and is identical to the static solution.
\noindent The resulting six dimensional metric now becomes,
\beq
\dis
ds^2 = \frac{\cosh^2(kz)}{\cosh^2(k\pi)}\,\frac{(b_0\,H_0)^2}{d^2}\,\sinh^2(-d\,|y|+d_0)\left[-dt^2\,+\,e^{2H_0 t}\,\delta_{ij}\,dx^i\,dx^j\right]\,+\,\frac{\cosh^2(kz)}{\cosh^2(k\pi)}\,b_0^2\,dy^2\,+\,c_0^2\,dz^2 \label{metric6}
\eeq
The above metric describes an inflationary model for the spatial three dimensions in a six-dimensional doubly warped spacetime.

\subsection{Pressures and Brane tensions on 3-branes}
Substituting eqn.(\ref{eq:9}), eqn.(\ref{eq:10}) and eqn.(\ref{eq:11}) in $G_{zz}$ component of Einstein's equation and integrating over an infinitesimal interval across the two boundaries at $y=0$ , $y=\pi$, we obtain respectively,\
\beq
\barr{rcl}
[-p_1(z)\,+\,\lambda_1(z)]|_{y=0} & = & 8M^4\,\dis\sqrt{\frac{-\Lambda_B}{10 M^4}}\,\coth (d_0)\,sech(kz)\\                         \label{eq:V_1}
[-p_2(z)\,+\,\lambda_2(z)]|_{y=\pi} & = & -8M^4\,\dis\sqrt{\frac{-\Lambda_B}{10 M^4}}\,\coth (-d\pi\,+\,d_0)\,sech(kz)                               
\earr
\eeq
We find the 4-branes located at $y=0$ and $y=\pi$ have z-dependent brane tensions. At $y=$ constant \,4-branes in the static limit when $ p_1\,=\, p_2=0 $, $d_0\longrightarrow \infty $ and $H_0\longrightarrow 0$  the brane tensions become equal and opposite.\\
Similarly, using eqn.(\ref{eq:9}), eqn.(\ref{eq:10}) and eqn.(\ref{eq:11}) in \,$G_{yy}$\, component of Einstein's equation and integrating over an infinitesimal interval across the two boundaries at $z=0$ , $z=\pi$ respectively we get,\\
\beq
\barr{rcl}
\dis
[-p_3(y)\,+\,\lambda_3(y)]|_{z=0} & =  & 0 \:,\qquad                                      \label{eq:V_3}
[-p_4(y)\,+\,\lambda_4(y)]|_{z=\pi} \, = \,  -\dis 8M^4\dis\sqrt{\frac{-\Lambda_B}{10 M^4}}\,\tanh (k\pi)                         
\earr
\eeq
\\
\noindent Now, each of the 3-branes lie at the intersection region of two 4-branes and their brane tensions, pressures and energy densities, to the leading order, are the algebraic sum of these quantities of two such 4-branes. The Standard model 3-brane that is located at $y=\pi$\,,\,$z=0$\,therefore has,\\
\beq
\dis
-p_{vis}\,+\,\lambda_{vis} = -8M^4\,\sqrt{\frac{-\Lambda_B}{10 M^4}}\,\coth (-d\pi\,+\,d_0)        \label{eq:vis}
\eeq

\noindent On the other hand, Planck brane is identified as the 3-brane situated at $y=0$\,,\,$z=\pi$. Hence,
\beq
\dis
-p_{pl}\,+\,\lambda_{pl} = 8M^4\,\sqrt{\frac{-\Lambda_B}{10 M^4}}\,\left[\coth(d_0)\,sech(k\pi)\,-\,\tanh(k\pi)\right]        \label{p_pl}
\eeq

\noindent Once again one can readily retrieve the brane tensions of the static six-dimensional doubly warped spacetime by going to the static limit : $d_0\longrightarrow \infty.$ and $p_{pl}=0$.

\noindent The combinations of pressures and brane tensions on the other two 3-branes located at $y=0$\,,\,$z=0$\, and at \,$y=\pi$\,,\,$z=\pi$\, can similarly be obtained as,\
\beq
-p(0,0)\,+\,\lambda(0,0)\,=\,\dis 8M^4\,\sqrt{\frac{-\Lambda_B}{10 M^4}}\,\coth(d_0)              \label{p(0,0)}
\eeq
\beq
-p(\pi,\pi)\,+\,\lambda(\pi,\pi)\,=\,\dis -8M^4\,\sqrt{\frac{-\Lambda_B}{10 M^4}}\,\left[\coth(-d\pi\,+\,d_0)\,sech(k\pi)\,+\,\tanh(k\pi)\right]                    \label{p_pi}
\eeq
\\

\subsection{Hubble parameter in 6D doubly warped spacetime}\
 We now find that on the $4$D world the Hubble parameter as a function of both the compact co-ordinates. Following \cite{kim}, by a suitable co-ordinate transformation eqn.(\ref{metric6}) can be recast as :
\beq
ds_{4}^2 = -dt^2\,+\,e^{2 H(y,z)t}\delta_{ij}\,dx^i\,dx^j
\eeq

\noindent where the effective 4D Hubble parameter is 
\beq
\barr{rcl}
H(y,z) = \dis \sqrt{\frac{-\Lambda_B}{10 M^4}}\,sech(kz)\,cosech(-d\, |y| + d_0)\\[1ex]  \label{Hubble}
\earr
\eeq

\noindent On the visible brane $(y\,=\,\pi\,,\,z\,=\,0)$ :
\beq
\dis
H(\pi,0)\,\equiv\,H_{vis}\, =\, \sqrt{\frac{-\Lambda_B}{10 M^4}}\,cosech\dis\left(-\frac{b_0 k \pi}{c_0 \cosh(k\pi)}\, + d_0\right)                        \label{Hubblev}
\eeq
On the Planck brane  $(y\,=\,0\,,\,z\,=\,\pi)$ :
\beq
\dis
H(0,\pi)\,\equiv\,H_{pl}\, =\, \sqrt{\frac{-\Lambda_B}{10 M^4}}\,sec(k\pi)\,cosech(d_0)
\eeq

\noindent However, the present universe is dominated by cosmological constant whose equation of state is :\
\beq
\rho_{vis} = -p_{vis} = \Lambda_{vis}
\eeq

\noindent Now, eqn. ({\ref{eq:vis}}) can also be written as,\
\beq
\dis
\lambda_{vis} +\,\Lambda_{vis}\,=-\,8M^4\,\sqrt{\frac{-\Lambda_B}{10 M^4}}\,\coth (-d\pi\,+\,d_0)               \label{l_vis}
\eeq
But eqn.({\ref{l_vis}}) can be written as 
\beq
 \lambda_{vis} =\,-\Lambda_{vis}\,-\,8M^4\,\sqrt{\frac{-\Lambda_B}{10 M^4}}\,\coth (-d\pi\,+\,d_0)              \label{l_vis_1}
\eeq
which shows Standard model brane possesses negative tension. After squaring eqn.({\ref{l_vis_1}}) we neglect $ \Lambda^2_{vis} $ and for $ \Lambda_B\,<\,0 $ we finally obtain
\beq
\Lambda_{vis}\,=\,\frac{1}{2}\left[\frac{\lambda^2_{vis}}{(8 M^4)\,\sqrt{\frac{-\Lambda_B}{10 M^4}}\,\coth (-d\pi\,+\,d_0) }\,-\,8 M^4\,\sqrt{\frac{-\Lambda_B}{10 M^4}}\,\coth (-d\pi\,+\,d_0)\right]               \label{Lambda}
\eeq
It may be noted that eqn.({\ref{Lambda}}) is the generalisation of eqn.({\ref{eq:cos_eff}}) extended to the non-static scenario of six-dimensional doubly warped spacetime.
\\
\noindent In the Randall Sundrum model the induced cosmological constant on the visible brane is zero thereby making it a flat brane which happens due to exact cancellation between bulk cosmological constant and corresponding brane tension. However, the presence of a negative pressure in our universe allows it to expand exponentially. As a result now the brane tension of the visible brane does not counter balance the contribution of the bulk cosmological constant exactly thus inducing a net positive cosmological constant on our universe. The shift in the value of brane tension from it's static value in the presence of additional matter is $ \delta \lambda_{vis} $ and therefore we write,
\beq
 \delta \lambda_{vis}\,=\, \lambda_{vis}\,-\,\lambda_{vis}|_{(static)}          \label{del_l}
\eeq

\noindent where,
\beq
\lambda_{vis}|_{(static)} \,=\,-8M^4\dis\sqrt{\frac{-\Lambda_B}{10 M^4}}              \label{eq:l_stat}
\eeq
\\
Then eqn.({\ref{l_vis_1}}) becomes
\beq
\lambda_{vis} =\,-\Lambda_{vis}\,+\,\lambda_{vis}|_{(static)}\dis(1\,+\,\varepsilon)
\eeq

\beq
\Rightarrow \delta \lambda_{vis}\,=\,-\Lambda_{vis}\,+\,\varepsilon\,\lambda_{vis}|_{(static)}
\eeq
where $ \varepsilon $ is a small positive quantity in excess of unity that appears only in the non-static case. Now squaring eqn.({\ref{l_vis}}) and using eqn.({\ref{del_l}}), eqn.({\ref{eq:l_stat}}) and further neglecting $ \delta \lambda^2_{vis} $ and $\Lambda_{vis}^2$ {\large[} as $\delta\lambda^2_{vis}$  and $ \Lambda_{vis}^2 $ are extremely small quantities {\large]} we obtain,
\beq
\delta \lambda_{vis}\,=\,\frac{(8 M^4)^2\,H_{vis}^2\,-\,2\,\Lambda_{vis}\,\lambda_{vis}|_{(static)}}{2[\lambda_{vis}|_{(static)}\,+\,\Lambda_{vis}]}          \label{eq:del_V}
\eeq
\\
\noindent\textbf{Estimation:} (in Planckian unit) \\

\noindent The present Hubble parameter on our universe is given by :
\beq
H_{vis}\,=\,1.233\,\times 10^{-61}M_{pl}
\eeq
and the present cosmological constant is known to be :
\beq
\Lambda_{vis}\,=\,10^{-124}M_{pl}^2
\eeq
Substituting all these values in eqn. (\ref{eq:del_V}) we find
\beq
 \delta \lambda_{vis} \,=\,- 1.923 \times 10^{-121}
\eeq
\noindent showing $ \delta \lambda_{vis} $ is a negative quantity and hence we must have $ \lambda_{vis}\,<\,\lambda_{{vis}|static} $ . This indicates that the static situation of the universe i,e. when eqn.({\ref{eq:12}}) is exactly valid, is highly unstable. With a shift from the static condition, a value as small as of the order of $ 10^{-121} $ is enough for the universe to switch into a non-static system that expands exponentially as the time passes by. 

\subsection{Length of extra dimensions}\
Let us now determine the size of extra dimensions \,$L_y$\,and \,$L_z$\, along the compact co-ordinates $y$ and $z$ respectively. The proper distance of the extra dimensions between $ y\,=\,0 $ to $ y\,=\,\pi $ and $ z\,=\,0 $ to $ z\,=\,\pi $ are
\beq
L_y  =  \dis\int_{0}^{\pi}\sqrt{g_{55}}\,dy =\, b(z) b_0 \pi                    \label{l_y}
\eeq
\beq 
L_z  =  \dis\int_{0}^{\pi}\sqrt{g_{66}}\,dz =\, c_0 \pi                         \label{l_z}
\eeq
We note that $L_y$ depends on the extent of warping of compact $z$ co-ordinate. This means an observer sitting at $z=0$ orbifolded fixed point will find \,$L_y$\, different from an observer present at 4-brane located at \,$z=\pi$ . This feature is a consequence of successive warpings of the spacetime. As the number of extra spatial dimensions increases, the proper lengths get warped even more except one. It would not arise if we had  considered independent warping of spacetime : $ [M^{1,3}\times S^1/Z_2\times S^1/Z_2] $. Let us now express \,$L_y$\, in terms of brane tensions and pressures. On the visible brane $b(z=0)=\dis\frac{1}{\cosh(k\pi)}$. Using eqn.({\ref{eq:vis}}) and eqn.({\ref{p(0,0)}}), the length of compact $y$ co-ordinate between the interval $0$ to $\pi$ turns out to be :
\beq
L_y = \dis \frac{b_0\pi}{\cosh(k\pi)}\, =\, \dis \frac{1}{2}\frac{1}{\dis\sqrt{\frac{-\Lambda_B}{10 M^4}}}\,
        \ln \left[\frac{\dis\left(\frac {-p_{vis}\,+\,\lambda_{vis}}{8M^4}\,+\,\sqrt{\frac{-\Lambda_B}{10 M^4}}\right)\,\dis\left(\frac{-p(0,0)\,+\,\lambda(0,0)}{8M^4}\,+\,\sqrt{\frac{-\Lambda_B}{10 M^4}}\right)}
             {\dis\left(\frac {-p_{vis}\,+\,\lambda_{vis}}{8M^4}\,-\,\sqrt{\frac{-\Lambda_B}{10 M^4}}\right)\,\dis\left(\frac{-p(0,0)\,+\,\lambda(0,0)}{8M^4}\,-\,\sqrt{\frac{-\Lambda_B}{10 M^4}}\right)}\right]
\eeq

\noindent By similar procedure $L_y$ measured from the Planck brane $b(z=0)=1$ is,
\beq
L_y = \dis b_0\pi\, =\,\dis \frac{1}{2}\frac{\cosh(k\pi)}{\dis\sqrt{\frac{-\Lambda_B} {10 M^4}}}\,
        \ln \left[\frac{\dis\left(\frac {-p_{vis}\,+\,\lambda_{vis}}{8M^4}\,+\,\sqrt{\frac{-\Lambda_B}{10 M^4}}\right)\,\dis\left(\frac{-p(0,0)\,+\,\lambda(0,0)}{8M^4}\,+\,\sqrt{\frac{-\Lambda_B}{10 M^4}}\right)}
             {\dis\left(\frac {-p_{vis}\,+\,\lambda_{vis}}{8M^4}\,-\,\sqrt{\frac{-\Lambda_B}{10 M^4}}\right)\,\dis\left(\frac{-p(0,0)\,+\,\lambda(0,0)}{8M^4}\,-\,\sqrt{\frac{-\Lambda_B}{10 M^4}}\right)}\right]    \label{eq:L_y_pl}
\eeq
\noindent $L_y$ determined from eqn. ({\ref{eq:L_y_pl}}) is the proper length of compactified $y$ co-ordinate as observed from the Planck brane. \\
It may be noted from eqn.({\ref{l_y}}) and eqn.({\ref{l_z}}) that $ L_y $ cannot be determined independently of $ L_z $. This feature is again a result of geometry of spacetime that is doubly warped in succession.
\\
\noindent Similarly, the size of the extra dimensions along the $z$ co-ordinate  can be expressed in terms of brane tensions and pressures by subtracting eqn.({\ref{p_pi}}) from eqn.({\ref{p_pl}}) and using eqn.({\ref{eq:vis}}) and eqn.({\ref{p(0,0)}}) :
\beq
L_z = c_0 \,\pi = \frac{1}{\dis\sqrt{\frac{-\Lambda_B}{10 M^4}}}\,sech^{-1}\dis\left[\frac{{-p_{pl}\,+\,p(\pi,\pi)}\,-\,\lambda(\pi,\pi)\,+\,\lambda_{pl}}{{-p(0,0)\,+\,p_{vis}}\,-\,\lambda_{vis}\,+\,\lambda(0,0)}\right]
\eeq

\subsection{Extent of warping on the visible brane}\
We examine the status of gauge hierarchy problem in our scenario. We consider the action of a free scalar propagating on the visible brane\\
\beq
S_{H} = \int d^4x \, \sqrt{-g_{vis}} \;
   \left[ g^{\mu\nu}_{vis} \, D_{\mu}H \, D_{\nu}H - m_0^2 \,
H^2\right] \ ,
\eeq 
the Planck scale mass $m_0$ is warped to 
\beq m = m_0 \,
\frac{b_0 H_0}{d } \sinh\left(\dis -\,\frac{b_0k\pi}{c_0 \cosh{k\pi}} + d_0\right) \dis\frac{1}{\cosh(k\pi)}\,e^{\frac{3}{4}H_0t} 
\eeq 
Near the static limit, it reduces to
\beq
m\,=\,m_0 \,  \frac{\exp(-d \pi)}{\cosh (k \pi)}\,e^{\frac{3}{4}H_0t}          \label{hier}
\eeq
 This ratio shows there cannot be equal warping in $y$ and $z$ directions simultaneously. If there is substantial warping along $y$- direction, in order to avoid a new undesirable hierarchy between $b_0$ and $c_0$, one must have small warping along $z$-direction and vice-versa. Further when we consider the present age of the universe, which is close to $ t = 13.4 \times 10^ 9 $ years then $ e^{\frac{3}{4}H_0t} = 2.06298 $. It may be noted that in the cosmological time scale $ e^{\frac{3}{4}H_0t} $  has negligibly small  variation. \\
\subsection{Estimation of length of extra dimension near static limit in doubly warped spacetime}
Since $ \dis\frac{m}{m_0}\approx 10^{-16} $, eqn.({\ref{hier}}) can be written as :
\beq
d\pi\,=\,\frac{b_0 k\pi}{c_0 \cosh(k\pi)}\,=\,-\ln\dis\left[\frac{\cosh(k\pi)\times 10^{-16}}{2.06298}\right]             \label{ratio}
\eeq
We have already mentioned that in order to avoid large hierarchy between the moduli $ b_0 $ and $ c_0 $, we require unequal warping along the 
two extra dimensions. For example we consider little warping along the 
compact $z$ co-ordinate a large warping along $y$ is necessary. In this situation we must have small value of $k$ such that $ k\leq 1 $ . 
The hierarchy between the moduli is minimum when $ k \sim 0.1 $ and from eqn.({\ref{ratio}}) this ratio is,
\begin{center}
\beq
 \dis\frac{b_0}{c_0}\,=\,125.419
\eeq
\end{center}
Also from eqn.({\ref{ratio}}),
\beq
d\,=\,11.948
\eeq
On visible brane $( y=\pi, z=0 )$ using eqn.({\ref{d}}), eqn.({\ref{l_y}}) and  eqn.({\ref{l_z}}) we can write, 
\beq
\frac{d}{k}\,=\,\frac{L_y}{L_z}
\eeq
and therefore for $ k=0.1 $
\beq
\frac{L_y}{L_z}\,=\,\frac{b_0}{c_0}\times sech(0.314)\,=\,119.480
\eeq
This implies $ L_y = 119.48\, L_z $ and $ b_0 = 125.419\, c_0 $
Thus there is a little hierarchy of order two between the to moduli. It may be noticed that a small hierarchy of order one also exists in the
original RS model due to the choice $k r \sim 11.5$. Just as the modulus is stabilised in such model by introducing a scalar field in the bulk
with appropriate vacuum expectation values (vev) at the two boundaries \cite{gw}, here also both the moduli can be stabilised by employing
two independent scalar fields in the bulk with absolute vev at the boundaries.

\subsection{Expressing Gauge hierarchy problem in terms of ratio of Hubble parameters} 
\noindent We have seen in section 2 that in five dimensions the non-zero Hubble parameter on the visible brane has been expressed as a fine tuning condition between visible brane tension $ \lambda_2 $ and bulk cosmological constant $ \Lambda_5 $. In the six dimensional doubly warped brane world scenario $ H_{vis} $  can be expressed in terms of fine tuning between $\lambda_{vis} $, $ p_{vis} $ and six-dimensional bulk cosmological constant $ \Lambda_B $. So we can write $ H_{vis} $ as,
\beq
H_{vis}\,=\,\sqrt{\dis\frac{(\lambda_{vis}\,+\,\Lambda_{vis})^2}{64 M^8}\,-\,\dis\left(\frac{-\Lambda_B}{10 M^4}\right)} \,= \,1.233 \times 10^{-61} M_{pl}            \label{24}
\eeq

\noindent The ratio of Planck brane to visible brane Hubble parameter is,
\beq
\frac{H_{pl}}{H_{vis}}\,=\,\frac{\sqrt{\dis\frac{(\lambda_{pl}\,-\,p_{pl})^2}{64 M^8}\,-\,\dis\left(\frac{-\Lambda_B}{10 M^4}\right)[1\,-\,2 \coth(d_0)sech(k\pi)\tanh(k\pi)]}}{\sqrt{\dis\frac{(\lambda_{vis}\,+\,\Lambda_{vis})^2}{64 M^8}\,-\,\dis\left(\frac{-\Lambda_B}{10 M^4}\right)}\,}             \label{ratio_1}
\eeq

\noindent Near the static limit where $ d_0 $ is large compared to $ -d\pi $  but $ \lambda_{vis},\, p_{pl},\, p(0,0) \neq 0 $  so that $\coth(d_0) $ is not exactly equal to $ 1 $. Then eqn.({\ref{p_pl}}) , eqn.({\ref{p(0,0)}}) and eqn.({\ref{l_vis}}) can be written as :
\beq
\dis
\lambda_{pl}\,-\,p_{pl} =\,-(\lambda_{vis}\, + \,\Lambda_{vis})\, \left[sech(k\pi)\,-\,\tanh(d_0)\tanh(k\pi)\right]    \label{21}
\eeq

\beq
\lambda(0,0)\,-\,p(0,0)\,=\,\dis 8M^4\,\sqrt{\frac{-\Lambda_B}{10 M^4}}\,\coth(d_0)\,=\,-(\lambda_{vis}\, + \,\Lambda_{vis})     \label{22}
\eeq

\beq
\lambda_{vis}\, + \,\Lambda_{vis}\,=\,-\,8M^4\,\sqrt{\frac{-\Lambda_B}{10 M^4}}\,\coth (d_0)      \label{23}
\eeq

\noindent Let $ \acute{k}\,=\,\dis\sqrt{\frac{-\Lambda_B}{10 M^4}} $ and using eqn.({\ref{21}}) , eqn.({\ref{22}}) and eqn.({\ref{23}}) we can express the fine-tuning of Higgs mass in terms of ratio of Hubble parameters of the Planck brane to that of the visible brane as follows:
\beq
 10^{-16} = \exp\dis(-\acute{k}L_y)\,sech(\acute{k}L_z)\,=
\,\dis\frac{\dis\sqrt{\left(\dis\frac{\lambda_{pl}\,-\,\dis p_{pl}}{8 M^4}\right)^2\,-\left(\dis\frac{-\Lambda_B}{10 M^4}\right)\,[sech(k\pi)\,-\,\tanh(d_0)\tanh(k\pi)]^2}}{\sqrt{\left(\dis\frac{\lambda_{vis}\,+\,\dis\Lambda_{vis}}{8 M^4}\right)^2\,-\left(\dis\frac{-\Lambda_B}{10 M^4}\right)}\,[sech(k\pi)\,-\,\tanh(d_0)\tanh(k\pi)]}\,sech(k\pi)           \label{ratio_2}
\eeq
\begin{center}
\beq
\simeq \dis\frac{H_{pl}}{H_{vis}}
\eeq
\end{center}
and using eqn.({\ref{24}})
\beq
\Rightarrow \dis H_{pl}\,\simeq\,10^{-77}M_{pl}
\eeq
It is to be noted that eqn.({\ref{ratio_1}}) and eqn.({\ref{ratio_2}}) are equivalent because for moderate value of $d_0$, $ \tanh(d_0) $ and $ \coth(d_0) $ are quite close and at the same time we have taken $ k \leq 1 $ to restrict hierarchy between the moduli. It is to be noted that this result closely resembles to the results obtained in $ 5 $D RS scenario \cite{kim}.  Similarly our result shows that resolution of fine tuning of Higgs mass requires a severe tuning on Planck brane between $ \lambda_{pl} $, $ p_{pl} $ and bulk cosmological constant $ \Lambda_B $ in six-dimensional doubly warped braneworld model.

\section{Matter in the bulk:}

Let us now assume that instead of brane the bulk contains pressure less matter. This generalises the case of putting as cosmological constant in the bulk. Then the six dimensional action becomes  
\beq
\barr{rcl}
 S & = & \dis S_6 + S_5 + S_4 \\[1ex]
 S_6 &  =  & \dis \int {d^4 x} \, {d y} \, {d z} \, \sqrt{-g_6} \;\left(\frac{M^4}{2}R_6 + {\cal L}_{Bulk}\right) 
\\[2ex]
 S_5 & = & \dis \int {d^4 x} \, {d y} \, {d z} \, \left\{\sqrt{-g_5}\;[\,{\cal L}_1\:{\delta(y)}\,+\,{\cal L}_2\, {\delta(y-{\pi})}\,]\,-\,\sqrt{-g_5}\;[ \,\lambda_1(z) \: {\delta(y)} + \lambda_2(z) \: {\delta( y - \pi)}\,] \,\right\}
\\[1.5ex]
 &  +   & \dis \int {d^4 x} \, {d y} \, {d z} \, \left\{\sqrt{-\tilde{g}_5}\;[\,{\acute{\cal L}}_3\:{\delta(z)}\,+\,{\acute{\cal L}}_4\:{\delta(z-{\pi})}\,]\,-\,\sqrt{-\tilde{g}_5}\ [ \,\lambda_3(y) \: {\delta(z)} + \lambda_4(y) \: {\delta( z - \pi)}\,]\right\}
\\[2ex]
 S_4 & =  & \dis \int d^4 x dy dz \sqrt{-g}[{\cal L} - \lambda] \ .
\earr
\eeq

On varying the action, we get Einstein's equation in this case as:

\beq
\barr{rcl}
- \dis M^4 \, \sqrt{-g_ 6} \, \left( R_{MN} - \, \frac{R}{2} \, g_{MN}\right)
& = & \dis (T_{Bulk})^{K}_{M}\,  g_{KN}\,\sqrt{-g_ 6} \: \\
 & + & \dis 
   
 \dis \left[\left(T^{\gamma}_{\beta}\,g_{\alpha\gamma}\right)_1\,+\,\,\dis \left(T^{\gamma}_{\beta}\,g_{\alpha\gamma}\right)_2\, \right] \sqrt{-g_5} \,\delta^{\alpha}_{M} \, \delta^{\beta}_{N}  \\[1ex] 
& + & \dis \left[\left(T^{\tilde{\gamma}}_{\tilde{\beta}}\,\widetilde g_{\tilde{\alpha} \tilde{\gamma}}\right)_3\,+\,\left(T^{\tilde{\gamma}}_{\tilde{\beta}}\,\widetilde g_{\tilde{\alpha} \tilde{\gamma}}\right)_4\right]  \sqrt{-\widetilde g_5}\,  \delta^{\tilde{\alpha}}_{M} \, \delta^{\tilde{\beta}}_{N} \\[1ex]
\earr
\eeq
\noindent where 
\beq
(T_{Bulk})^K_M\,=\,diag\dis\left(-\frac{1}{v^3(t)},0,0,0,0,0\right)
\eeq
and the energy momentum tensor on the 4-branes are given as before by eqn.({\ref{T_1}}) , eqn.({\ref{T_2}}) , eqn.({\ref{T_3}}) and eqn.({\ref{T_4}}) respectively.

\noindent In such a scenario, from $ (ii) $ and $ (tt) $  components of Einstein's equation one obtains,
\beq
\barr{rcl}
\, \dis \left(\frac{\dot v^2}{v^2}\,-\,\frac{\ddot{v}}{v}\right)v^3&=&\dis \frac{b^2\,a^2}{2M^4}             \label{eq:30}
\earr
\eeq

\noindent The 3-brane on which our FRW universe exists, resides at the intersection region of two 4-branes located at $y=\pi\,,\,z=0$. On this 3-brane the values of the warp factors $a(y=\pi)$and $b(z=0)$ are constant. Since the branes are delta-function sources, these values are unique irrespective of the type of its matter content and hence right hand side of eqn.({\ref{eq:30}}) is a constant quantity which implies left hand side must also be constant. To realise this let us assume that the FRW universe situated on the Standard model brane is described by $\Lambda$CDM model. As a result, on the visible brane we can take\
\beq
v(t)=v_0\sinh^m(H_0\,t)
\eeq

\noindent $v_0$ being a constant. From the solution of v(t), we can infer for $H_0t<<1$,\, $v(t)\longrightarrow(H_0\,t)^m$\, implying our universe was dominated by pressure less matter [for $ m = \frac{2}{3} $] in the past and when \,$H_0 t>>1$, \,$v(t)\longrightarrow\exp(H_0 t)$ shows at later times universe will be vacuum energy dominated .\\
\noindent Finally, on substituting the solution of $v(t)$ in LHS of eqn. (\ref{eq:30}) , 
we obtain $\dis\left(v_0^3\,m\,H_0^2\,\sinh^{3m-2}(H_0 t)\right)$ which is a constant quantity and is equal to $\dis\left(\frac{2}{3}\,v_0^3\,H_0^2\right)$\, only if $m=\frac{2}{3}$. 

\section{Conclusions :}
In this work we have addressed the issue of Hubble expansion in a six dimensional doubly warped braneworld model.
We have shown that 
as long as the bulk contains only cosmological constant and no other matter fields, the solution of the time dependent part of the 
scale factor $ v(t) $ is given by eqn.({\ref{eq:5}}) . This result is quite generic and independent of the 
number of warped extra dimensions.
Thus the exponential expansion phase of our Universe is a direct consequence of existence of a
bulk negative cosmological constant in a warped braneworld scenario. In such scenario we have 
derived the exact solution for the cosmological spacetime metric. 
Through out the analysis the branes are chosen as delta function sources.\\ 

We then extended our analysis with an additional pressure less matter in the bulk and found that the  
standard model brane metric has a solution for $ v(t) $ in the form $ v(t)\sim \sinh^m(H_0 t) $. This solution produces a deceleration expansion for 
small $ t $ and an accelerated one for large $ t $ for the visible Universe. This matches very well  with the observations.

In a multiply warped one needs an  unequal warping along the two compact extra dimensions $y$ and $z$ in order to restrict a large hierarchy between the 
respective moduli $ b_0 $ and $ c_0 $ \cite{Multiply}. The hierarchy is minimum for $ k=0.1 $ which produces a small warping along $z$ direction and 
quite large wrping along the compact $y$ direction. 
Under this situation we have shown that length $ L_y $ becomes elongated almost $120$ times more than $ L_z $.
Also the ratio of $L_y$ and $L_z$ is related the ratio of the moduli of the extra dimensions and hence $ b_0\approx 125 c_0 $.
 
From eqn.(\ref{Hubblev}) we find that Hubble parameter on our Universe is controlled by both the moduli $b_0$ and $c_0$ of the compact 
extra dimensional co-ordinates $y$, $z$ respectively. The factor $sech(k\pi)$ ( for the optimal choice $k \sim 0.1$ ) is very small 
and can be viewed as a small
perturbation  on the five dimensional braneworld model. Furthermore, $ H_{vis} $ is found to depend on lengths of both the  extra dimensions. 
 
The static limit of our work correctly reproduces RS metric (given by eqn.({\ref{metric}})) when the effective cosmological constant on the 3-brane
is zero giving rise to flat $3$-branes. However, if we introduce additional matter on the $4$-branes in the form of perfect fluid 
characterised by energy density $\rho$ and pressure $p$, the static RS model becomes unstable. Furthermore a negative 
fluid pressure (strong energy condition is no longer valid now ) on the visible brane initiates an exponential expansion of the Universe with time. 
Consequently, the visible brane tension now departs from its static value. In the present work we have shown that a very small shift is required for the 
Universe to undergo inflation and found it to be of the order of $ 10^{-121} $. Under this situation, $ H_{vis} $ is non-zero and a 
fine tuning between bulk cosmological constant $ \Lambda_B $ and visible brane tension $ \lambda_{vis} $ may give rise to a net cosmological constant 
on the Universe which is consistent with the present estimation.\\
It may also be noted that eqn.({\ref{Lambda}}) is a generalisation of eqn.({\ref{eq:cos_eff}}) of the work of \cite{shiromizu} in the non-static case
which essentially relates the induced cosmological constant on the brane with the bulk cosmological constant and the brane tension. 
We finally infer that although from ({\ref{hier}}) Higgs mass appears to be a time dependent quantity, however in the cosmological time scale its 
variation is negligible as shown in section 5.5. Our work thus brings out the dependence of cosmological expansion and related features in the
context of larger number of warped extra dimensional model.

\end{document}